%% file: paper.tex
\documentclass[5p]{elsarticle}

\usepackage[utf8]{inputenc}
\usepackage{amsmath,amssymb}
\usepackage[english]{babel}
\usepackage[table]{xcolor}
\definecolor{Blue}{rgb}{0.3,0.3,0.9}
\usepackage{graphicx}
\usepackage{dcolumn}
\usepackage{bm}
\usepackage{hyperref}
\hypersetup{
 colorlinks=true,
    linkcolor=blue,
    citecolor=blue,
    urlcolor=blue,
}
\usepackage{mdframed}
\usepackage{todonotes}
\graphicspath{{Figures/}}
\usepackage{comment}

\usepackage{booktabs}        
\usepackage{threeparttable}  
\usepackage{siunitx}         
\usepackage{colortbl}

\newcommand{\COMMENT}[1]{}

\journal{Nuclear Materials and Energy}
\begin{document}

\begin{frontmatter}

\title{Cycle-Consistent and Uncertainty-Aware Neural Surrogates for Tokamak Edge Plasmas}

\author[ornl]{A. Diaw\corref{cor1}}
\ead{diawa@ornl.gov}
\author[ornl]{S. De Pascuale}
\author[ornl]{J.-S. Park}
\author[ornl]{I. Paradela Perez}
\author[ornl]{J.D. Lore}
\author[differ]{S. Dasbach}

\cortext[cor1]{Corresponding author}
\address[ornl]{Fusion Energy Division, Oak Ridge National Laboratory, Oak Ridge, TN 37932, USA}
\address[differ]{DIFFER - Dutch Institute for Fundamental Energy Research, De Zaale 20, 5612 AJ Eindhoven, the Netherlands}

\fntext[notice]{This manuscript has been authored in part by UT-Battelle, LLC, under contract DE-AC05-00OR22725 with the US Department of Energy (DOE). The publisher acknowledges the US government license to provide public access under the DOE Public Access Plan (http://energy.gov/downloads/doe-public-access-plan).}

\begin{abstract}
The boundary and divertor plasma play a key role in how a tokamak removes power and particles. They set the heat fluxes, temperatures, densities, and the start of detachment. Predicting these values accurately is crucial for safely running current and future devices. However, detailed edge simulations that resolve these parameters are too slow for tasks such as parameter scans, optimization, or real-time control. To address this, machine learning surrogates are now often used instead of traditional edge-plasma simulations. Still, most standard models can only predict forward and cannot recover input parameters from observed data or check how reliable their predictions are. In this study, we introduce a cycle-consistent neural surrogate for edge plasmas. It combines a conditional U-Net forward model with an optimization-based inverse method that uses the frozen forward network. The forward model takes five control parameters and predicts two-dimensional plasma-state fields on the SOLPS-ITER mesh. The inverse method ensures consistency between forward and inverse predictions, offering a self-supervised quality check that does not need ground-truth labels during inference. We also train a group of multilayer perceptrons to predict electron temperature and density profiles at the outboard midplane and divertor targets, with uncertainty estimates. The variation among the committee members provides a reliability measure for real-time control and helps spot areas where more simulations are needed. The forward model achieves normalized root-mean-square errors below $2.6\%$ and Pearson correlations above $0.95$ for all plasma-state fields. Adding cycle-consistency regularization raises the average cyclical $R^2$ from $0.59$ to $0.99$ without reducing forward accuracy, and it allows recovery of the core fueling rate $\Gamma_{\mathrm{core}}$, which is hard to determine with forward-only training. The inverse method recovers all five control parameters with Pearson $r\ge 0.97$. Using a $k$-d tree warm start to build the database yields a completion rate above $95\%$, whereas in a comparable cold-started ensemble roughly $30\%$ of runs failed outright and a further fraction never reached steady state. With about $4\times10^6$ parameters, the model can generate full two-dimensional predictions in milliseconds, which is five to six orders of magnitude faster than the original SOLPS-ITER runs. This speed is enough for real-time control, thorough parameter scans, uncertainty analysis, and digital-twin applications.

\end{abstract}


\end{frontmatter}

\section{\label{sec:intro} Introduction}

Surrogate models have long been a workhorse of computational physics, standing in for first-principles solvers when direct numerical simulation is too slow to explore or optimize a problem at scale \cite{Wigley:2016, Scheinker:2015, Noack:2019, Diaw2024}. They are most useful precisely where simulation is hardest: multiscale physics, stiff governing equations, and the need for large parameter scans or real-time inference. But this is also where a surrogate is most likely to fail. A surrogate is only trustworthy if it reproduces the system where its behavior is sharpest, near thresholds, bifurcations, and steep spatial gradients; these are exactly the regions that are hardest to learn.

Scrape-off-layer (SOL) plasma is a textbook case. In the SOL, plasma--neutral and plasma--material interactions are set by kinetic processes, detailed atomic and molecular rates, and sheath physics acting on short spatial and temporal scales. Yet the quantities that matter for divertor design, scenario optimization, and feedback control, namely the target heat fluxes, the density profiles, and the radiation patterns, live at the macroscopic scale of the whole boundary and evolve over milliseconds to seconds. Predictively bridging these scales is essential for current and future devices, yet it remains computationally challenging~\cite{ref:solps_review, ref:edge_overview}.
  
For SOL physics, the state-of-the-art tool is SOLPS-ITER, which couples a two-dimensional Braginskii fluid solver (B2.5) to a kinetic Monte Carlo neutral transport model (EIRENE) on a triangulated vessel mesh~\cite{ref:eirene_manual}. This multiphysics code has been used extensively to design and interpret experiments on existing devices and to evaluate divertor concepts for future machines. It is also computationally expensive. Each run solves a stiff nonlinear fixed-point problem whose wall-clock time often exceeds several hours, whose convergence is highly sensitive to initialization, and whose failure rate climbs sharply when one scans broad parameter sets or enters detached regimes. For these reasons, exhaustive sweeps over input power, gas puff, and transport coefficients are computationally prohibitive, and the solver cannot be placed directly inside an optimization loop or a real-time workflow.

Data-driven surrogates for edge plasmas have received growing attention, using reduced models or neural networks trained on high-fidelity simulations to interpolate across parameter space~\cite{ref:ml_edge_2}. These surrogates already recover the essential features, including detachment, target fluxes, and profile shapes. Two limitations persist, however. Most are confined to one-dimensional profiles at a handful of fixed locations, and they rarely enforce any physical consistency between the forward and inverse mappings. As a result, a prediction arrives with no built-in way to tell whether it can be trusted, which is exactly what one needs in the sharp, undersampled regimes where the surrogate is most likely to be wrong.

We present here a cycle-consistent neural surrogate for two-dimensional edge-plasma fields. At its core is a conditional U-Net forward model $F_\theta$ that maps five scalar control parameters to the two-dimensional plasma state ($T_e$, $T_i$, $n_e$, $u_a$). A gradient-based inverse procedure then recovers the control parameters from a target field by optimizing through the frozen forward model. A cycle-consistency constraint ties the two together: parameters recovered from a field, when pushed back through $F_\theta$, must reproduce that field. This round-trip supplies a self-supervised quality metric that requires no ground-truth labels at inference time.

We choose a deterministic, parameter-conditioned U-Net. Edge-plasma databases are data-scarce, with a few hundred converged runs at best. In this regime, it is prudent to learn a single-valued operator from control parameters to fields rather than a full distribution. Unlike generative approaches such as variational autoencoders or autoregressive transformers, the U-Net carries no sampling variance and does not have the large data appetite of those models~\cite{Kingma_2019,Unet}. Our choice is pragmatic, and we believe it is well motivated. The resulting model has roughly 4.3 million parameters and returns a full two-dimensional prediction within milliseconds, fast enough for rapid scenario evaluation and for deployment in real-time plasma control~\cite{Degrave2022,Seo2024} and digital-twin workflows~\cite{Tang2024,Willcox2024}.

Alongside this two-dimensional model, we train a query-by-committee ensemble of multilayer perceptrons for the one-dimensional $T_e$ and $n_e$ profiles at the outboard midplane and the divertor targets. We restrict it to $T_e$ and $n_e$ as an initial demonstration of the method, and because these are the quantities measured directly by Thomson scattering, facilitating future validation. The ensemble supplies predictive uncertainty estimates for closed-loop control, and its committee variance flags the under-sampled regimes where new SOLPS-ITER runs are most needed, a query-by-committee strategy we have used in earlier work~\cite{DiawPRE20,DiawNME2022}. Together, the two surrogates cover the two deployment modes we care about: design and inverse analysis on one side, control and active-learning acquisition on the other.

The structure of this paper is as follows. Section~\ref{sec:database} details the simulation database and the k-d tree warm-start strategy. Section~\ref{sec:surrogate} describes the surrogate architecture, including the conditional U-Net forward model, the inverse inference procedure, and the cycle-consistency framework. Section~\ref{sec:results} evaluates forward model accuracy, inverse parameter inference, and cycle-consistency metrics on held-out test cases. Section~\ref{sec:profile_ensemble} presents the uncertainty-quantified one-dimensional profile ensemble for control and active learning. Section~\ref{sec:discussion} interprets the results, and Section~\ref{sec:conclusions} summarizes the conclusions and outlines future directions.

\section{Dataset}
\label{sec:database}

The training dataset was generated using SOLPS-ITER~\cite{Wiesen2015,Reiter2005} simulations of a deuterium plasma in DIII-D, employing the same lower-single-null configuration as used in Ref.~\cite{Lore2023}. This setup was developed using DIII-D discharge 174310 at $t=3500\,$ms, has been demonstrated to have excellent numerical stability, and has been used to assess divertor conditions and neutral dynamics~\cite{Lore2023,Park2024,Lasa2024}. Consequently, it serves as a convenient baseline for constructing surrogate models of the coupled plasma--neutral system and for comparison with other SOL machine-learning studies~\cite{Dasbach2023,Wiesen2024,Zhu2022,Zhu2025}. 

To span a representative operational space, we sample the input parameters using Latin hypercube sampling (LHS) over the bounds in Table~\ref{tab:solps_iter_sampling_ranges}. Each sample corresponds to a five-dimensional input vector:
\[
\mathbf{c} = (P_{\text{tot}}, \Gamma_{\text{core}}, \Gamma_{\mathrm{D}_2}, D_\perp, \chi_i),
\]
which sets the total power across the separatrix, the target core flux, the deuterium puff rate, and the uniform cross-field particle and heat transport coefficients. 

While LHS provides good space-filling coverage at coarse resolution, each simulation run effectively solves a stiff nonlinear fixed-point problem whose convergence is highly sensitive to the initial condition. In practice, the chosen restart file strongly influences the solver's convergence trajectory: depending on the location in parameter space, runs may require many iterations, stall, or fail to converge numerically. Nontrivial failure rates have been reported in previous work. In the large cold-started (fluid-neutral) ensemble of Dasbach and Wiesen~\cite{Dasbach2023}, about $29\%$ of runs ($1198$ of $4096$) diverged outright, and a further fraction of the surviving runs never reached steady state within the allotted runtime.

\begin{table}[!b]
\centering
\caption{Control parameters and their sampling ranges for the SOLPS-ITER database.}
\label{tab:solps_iter_sampling_ranges}
\resizebox{\linewidth}{!}{%
\begin{tabular}{l c c c c c}
\toprule
 & $P_{\text{tot}}$ & $\Gamma_{\text{core}}$ & $\Gamma_{\text{D}_2}$ & $D_{\perp}$ & $\chi_i$ \\
 & (MW) & (at/s) & (at/s) & (m$^2$/s) & (m$^2$/s) \\
\midrule
\rowcolor{blue!5}
\textbf{min} & 2 & $1 \times 10^{20}$ & $1.3 \times 10^{20}$ & 0.1 & 0.1 \\
\textbf{max} & 16 & $7.5 \times 10^{20}$ & $5 \times 10^{21}$ & 2.0 & 2.0 \\
\bottomrule
\end{tabular}%
}
\end{table}

To mitigate these issues, we adopt a branching $k$-d tree strategy~\cite{Bentley1975} that both partitions the input space and \emph{warm-starts} each simulation from the nearest already-converged design point, propagating reliable initial conditions outward from a single seed run. This raises the completion rate above $95\%$, compared with the roughly $70\%$ completion of the cold-started ensemble above. The construction and distance metric are detailed in Appendix~\ref{sec:appendix_warmstart}.

The detailed convergence detection procedure, including autocorrelation analysis and block averaging, is described in Appendix~\ref{sec:appendix_convergence}. The completed database comprises $762$ converged runs, of which two were discarded during quality screening, leaving $760$ for surrogate modeling. We split these $80/20$ into a training/validation set of $608$ runs and a held-out test set of $152$ runs. The split is performed at the level of whole simulations, so all mesh cells of a given run fall entirely within either the training or the held-out set, preventing information leakage. All the results shown in this paper are on the test set.

\section{Surrogate Model Design}
\label{sec:surrogate}

\begin{figure*} 
  \centering
  \includegraphics[width=\linewidth]{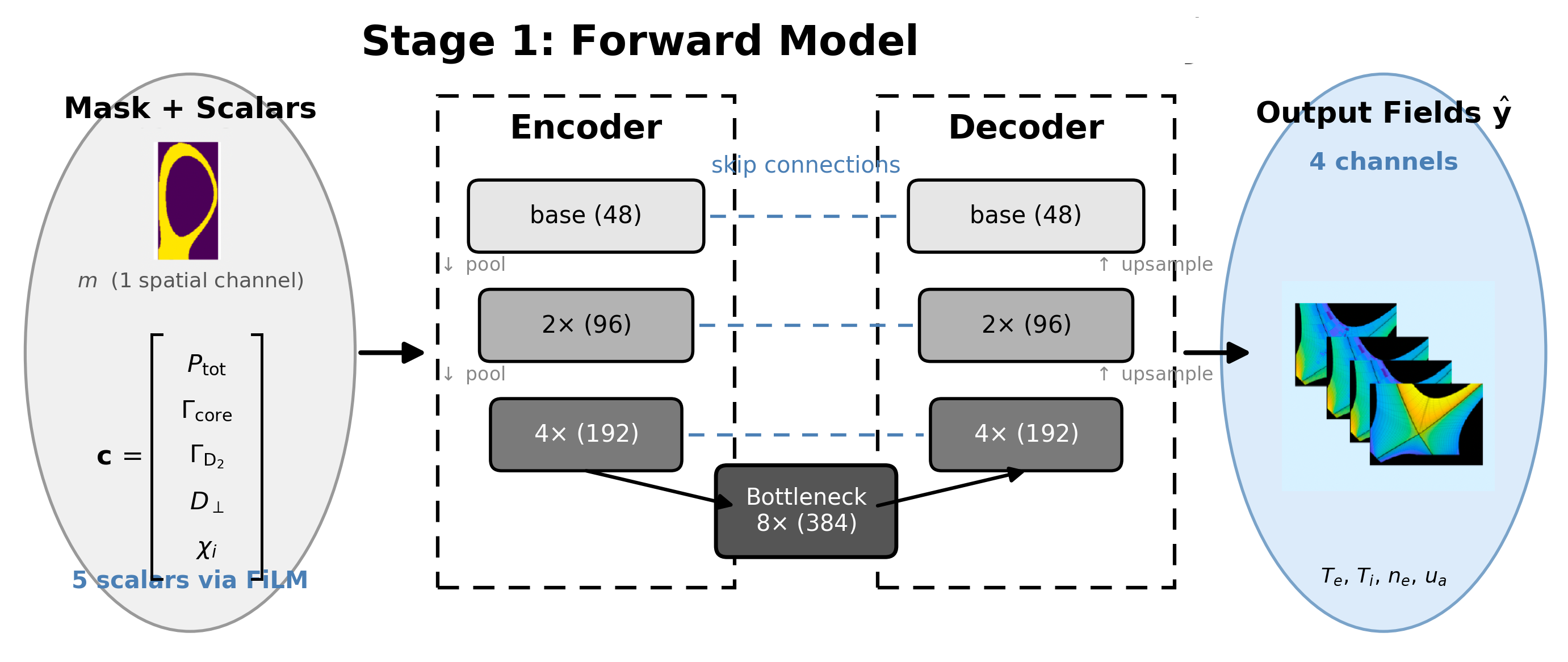}\\[3pt]
  \includegraphics[width=\linewidth]{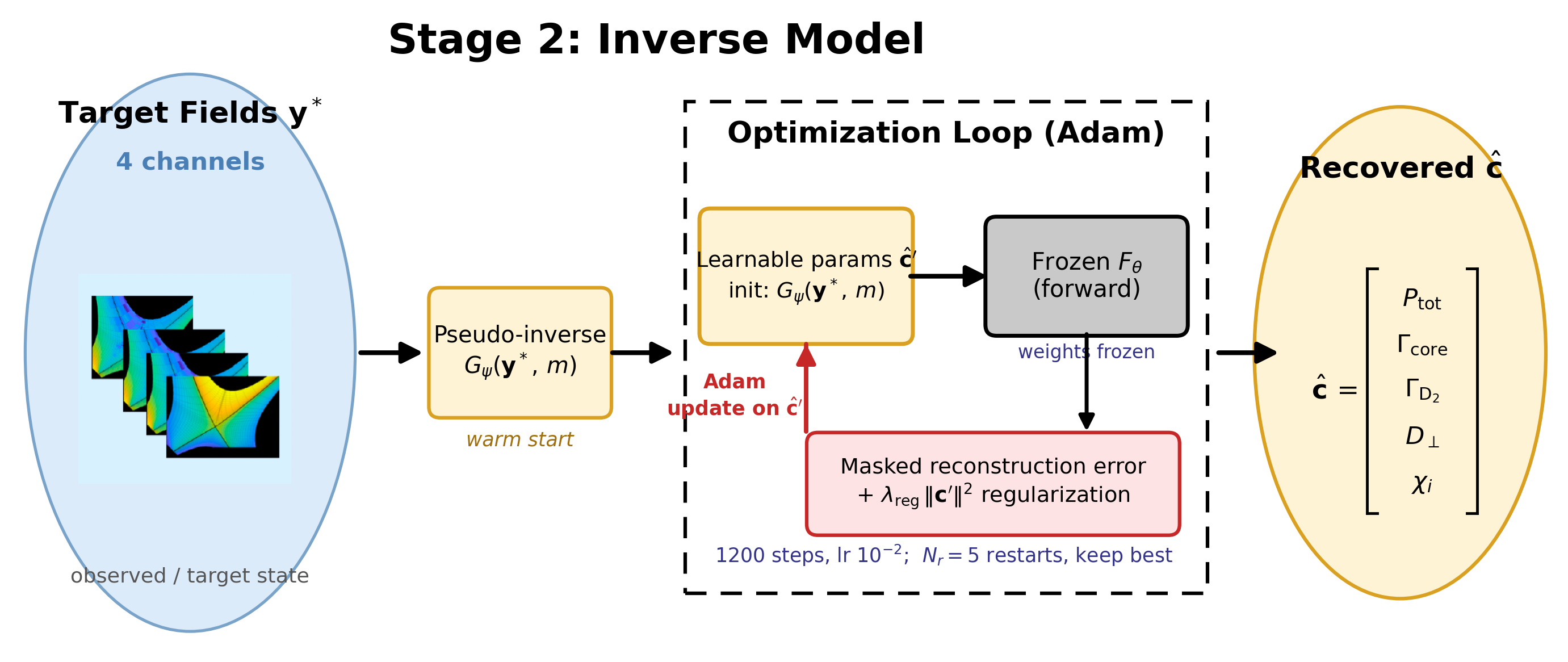}\\[3pt]
  \includegraphics[width=\linewidth]{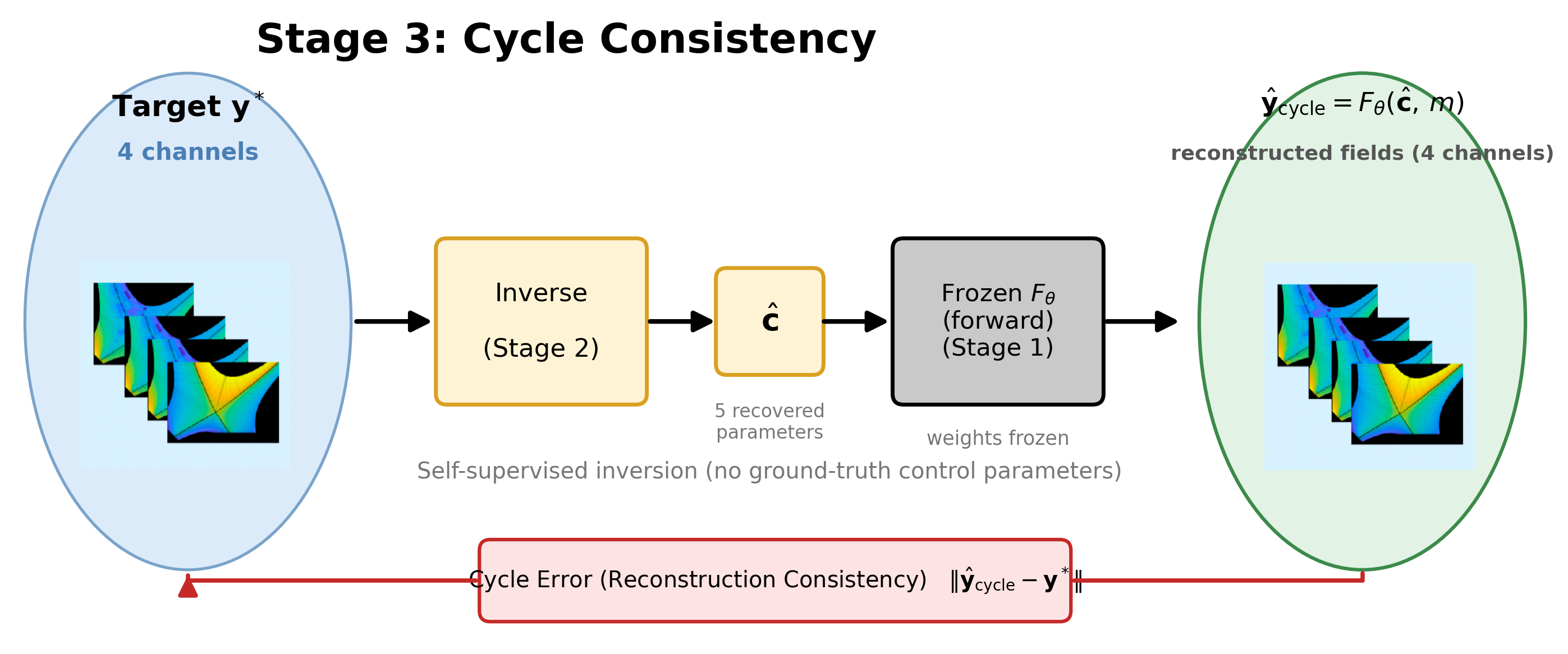}
  \caption{Surrogate model architecture: the three-stage inference cycle.
    \emph{(a)~Stage~1 (forward model)}: a conditional U-Net maps a single-channel geometry mask, with the five scalar control parameters injected via FiLM, to the four-channel plasma state ($T_e, T_i, n_e, u_a$); the fields are outputs of the model, not inputs.
    \emph{(b)~Stage~2 (inverse model)}: the control parameters are recovered from a target field by optimizing through the frozen forward model, warm-started by the learned pseudo-inverse $G_\psi$ (Fig.~\ref{fig:macc}) evaluated on the observed fields.
    \emph{(c)~Stage~3 (cycle consistency)}: the recovered parameters are passed back through the frozen model; the cycle loss validates that the round-trip preserves the target field, a self-supervised metric requiring no ground-truth parameters.
    The forward model $F_\theta$ and pseudo-inverse $G_\psi$ used here are produced beforehand by the training-time regularizer of Fig.~\ref{fig:macc}.}
  \label{fig:unet:model}
\end{figure*}

\begin{figure*}[!t]
  \centering
  \includegraphics[width=0.82\linewidth]{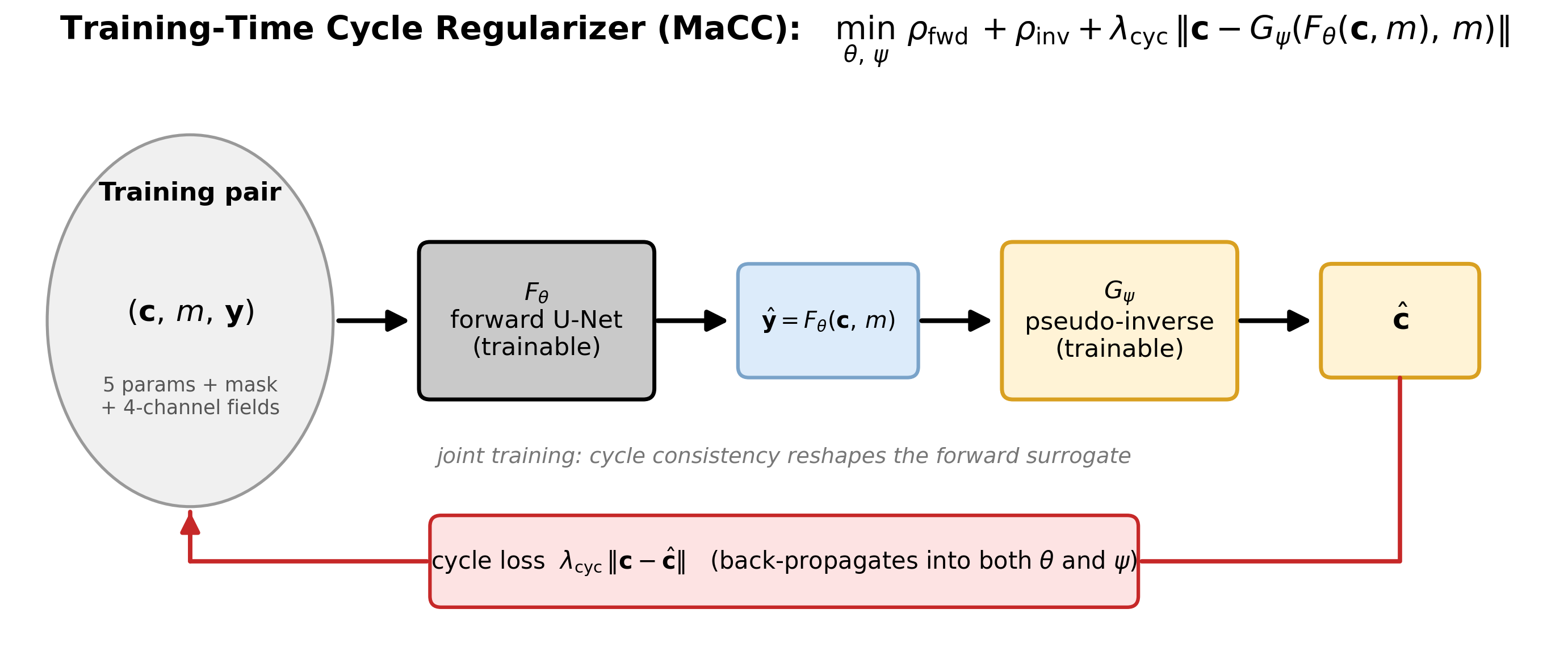}
  \caption{Training-time cycle-consistency regularizer. The forward model $F_\theta$ and the learned pseudo-inverse $G_\psi$ are trained jointly; the cycle penalty $\lambda_{\mathrm{cyc}}\|\mathbf{c}-G_\psi(F_\theta(\mathbf{c},m),m)\|$ back-propagates into both networks, so consistency reshapes the forward surrogate. Setting $\lambda_{\mathrm{cyc}}=0$ recovers the forward-only baseline; the weight is swept in the ablation of Sec.~\ref{sec:results_ablation}.}
  \label{fig:macc}
\end{figure*}

The surrogate framework follows the manifold and cycle-consistency paradigm. At inference, it runs as a three-stage cycle (Fig.~\ref{fig:unet:model}): a forward model, an optimization-based inverse, and a cycle-consistency self-check. All three stages share a single trained forward model $F_\theta$ and a learned pseudo-inverse $G_\psi$, which are produced beforehand by a training-time cycle-consistency regularizer (Sec.~\ref{sec:macc}, Fig.~\ref{fig:macc}). The workflow has three components:  (1) a conditional U-Net $F_\theta$ is trained on the SOLPS-ITER database to map scalar control parameters  $\mathbf{c}=(P_{\mathrm{tot}},\,\Gamma_{\mathrm{core}},\, \Gamma_{\mathrm{D}_2},\,D_\perp,\,\chi_i)$ and a spatial binary mask $m$ to the four-channel two-dimensional plasma state. Once trained, $F_\theta$ defines the learned physics manifold: every output $\hat{\mathbf{y}}=F_\theta(\mathbf{c},m)$ is, by construction, a physically plausible edge-plasma state, (2) an inverse model: given a target field $\mathbf{y}^*$, an optimization-based inverse procedure $G^\star$ recovers the control parameters $\hat{\mathbf{c}}$ by minimizing the discrepancy $\|F_\theta(\hat{\mathbf{c}},m)-\mathbf{y}^*\|$ through the frozen forward model via gradient descent, and (3) a cycle-consistency stage in which the recovered parameters are passed back through $F_\theta$ to obtain a reconstructed field $\hat{\mathbf{y}}_{\mathrm{cycle}} =F_\theta\!\bigl(G^\star(\mathbf{y}^*),m\bigr)$. The cycle-consistency loss $\|\hat{\mathbf{y}}_{\mathrm{cycle}}-\mathbf{y}^*\|$ measures whether the round-trip $\mathbf{y}^*\!\to\!\hat{\mathbf{c}}\!\to\! \hat{\mathbf{y}}_{\mathrm{cycle}}$ preserves the target, providing a self-supervised quality metric that does not require ground-truth parameter labels at test time.
        
\noindent
Each stage is described in detail in the following subsections.

\subsection{Conditional U-Net Forward Model}
\label{sec:forward}
The forward surrogate $F_\theta$ maps scalar control parameters $\mathbf{c}$ and the spatial binary mask $m$ to the multi-channel plasma state
\begin{eqnarray}
      \hat{\mathbf{y}} &=& F_\theta(\mathbf{c},\,m),\\
        \hat{\mathbf{y}}
    &=& \{T_e,\,T_i,\,n_e,\,u_a\},
\end{eqnarray}
on the SOLPS $(R,Z)$ grid, where $T_e$ and $T_i$ are the electron and ion temperatures, $n_e$ is the electron density, and $u_a$ is the parallel velocity of the deuterium ions. Although the SOLPS state arrays are dense in index space $(i,j)$, the physical plasma domain is irregular in $(R, Z)$. It occupies only a subset of the rectangular tensor used by convolutional neural networks (CNNs)~\cite{krizhevsky2012imagenet,lecun1989backpropagation}. Standard U-Net convolutions operate on all grid locations, so without masking, cells outside the valid plasma region would still contribute to normalization and loss, potentially biasing training toward non-physical/background regions. The binary mask, therefore, acts as a domain-of-validity selector: optimization and evaluation are restricted to physically meaningful mesh cells while preserving compatibility with efficient regular-grid
convolutions.

The scalar parameters $\mathbf{c}$ are standardized using the mean and standard deviation of the training set and injected into the network via Feature-wise linear modulation (FiLM)~\cite{Perez2018}: at each encoder and decoder level, the parameter vector is mapped through a learned affine layer that produces the per-channel scale and shift coefficients applied after group normalization. This yields a single-channel spatial input (the binary mask alone) while enabling the network to learn a parametric nonlinear operator across multiple plasma regimes.

Each output channel is standardized independently, with statistics computed on the training split over valid mesh cells only. Because the temperatures and density are positive and span several orders of magnitude, they are log-transformed before being centered and scaled to unit variance (with a small offset, $10^{-2}$ for $T_e,T_i$ and $10^{16}\,\mathrm{m^{-3}}$ for $n_e$, to regularize near-zero values). The parallel velocity $u_a$ changes sign along the flow and is instead standardized with a symmetric-log transform before centering and scaling. The five input scalars are standardized with a plain mean/standard-deviation z-score.

{\bf Encoder.}
The network follows a U-Net architecture~\cite{Unet} with three
resolution levels. At each level, an input feature map is processed by two $3\times3$ convolutions, each followed by group normalization and a SiLU activation:
\begin{eqnarray}
      \mathbf{Y}' & \;=\;  & \sigma\!\big(\mathrm{GN}(\mathbf{W}*\mathbf{Y})\big),\\
      \sigma(s) &= & s\cdot\mathrm{sigmoid}(s).
\end{eqnarray}
Downsampling is performed using $2\times2$ max pooling with a stride of 2; at each step, the spatial size halves while the number of channels doubles. The bottleneck applies the same double-convolution block at $8\times$ base channels.

{\bf Decoder.}
Each up-step uses a $2\times2$ transposed convolution to increase
spatial resolution and reduce channels, concatenates the corresponding
encoder feature map via a skip connection, and then applies two
$3\times3$ convolutions with group normalization and SiLU. After the last decoder block, a $1\times1$ convolution maps the base
channels to $C_{\mathrm{out}}=4$ output channels corresponding to the four plasma state fields. Because the targets are continuous, we use a linear activation at the final layer. Additionally, because $n_e \equiv n_i$ in the plasma, we omit the ion density channel.

For thermodynamic consistency, we enforce non-negativity of the temperature and density outputs ($T_e, T_i, n_e\ge0$) by clamping negative values during post-processing.

\subsection{Inverse Model}
\label{sec:inverse}

The inverse problem, in which we infer control parameters $\mathbf{c}$ from target plasma fields $\mathbf{y}^*$, is ill-posed: many parameter combinations can yield similar field configurations. Rather than training a separate inverse network, we adopt an optimization-based approach that leverages the differentiability of the frozen forward model $F_\theta$. We denote this optimization-based inverse $G^\star$; unlike $F_\theta$ and the pseudo-inverse $G_\psi$ (Sec.~\ref{sec:macc}), it has no trainable parameters of its own: it recovers $\hat{\mathbf{c}}$ by optimizing the control parameters directly through the frozen forward model.

Given a target field $\mathbf{y}^*$ and spatial mask $m$, we solve for the parameters $\hat{\mathbf{c}}$ that minimize the discrepancy between the forward prediction and the target:
\begin{eqnarray}
      \hat{\mathbf{c}}
    &=& \arg\min_{\mathbf{c}'}\;\mathcal{L}_{\text{inv}}(\mathbf{c}'),
      \nonumber\\
\mathcal{L}_{\text{inv}}(\mathbf{c}')
   & = & \frac{\displaystyle
        \sum_{c,i,j} w_c\,m_{ij}
        \bigl(F_\theta(\mathbf{c}',m)_{c,i,j}-y^*_{c,i,j}\bigr)^{\!2}}
      {\displaystyle\sum_{c,i,j} w_c\,m_{ij}} \nonumber \\
    && + \lambda_{\text{reg}}\|\mathbf{c}'\|^2,
      \label{eq:inverse}
\end{eqnarray}
where $w_c$ are the per-channel weights and $\lambda_{\text{reg}}$ is an $L_2$ regularization coefficient that prevents the recovered parameters from drifting to unphysical extremes. Optimization is performed using Adam~\cite{adam} with learning rate $10^{-2}$ for $1200$ steps through the frozen forward model, with gradient-based parameter updates computed via automatic differentiation~\cite{PyTorch}. The optimization is warm-started from the pseudo-inverse estimate $\hat{\mathbf{c}}_0=G_\psi(\mathbf{y}^*,m)$ (Sec.~\ref{sec:macc}), which is computed from the observed fields alone. To mitigate local minima, we employ $N_r=5$ random restarts about this estimate, perturbed with Gaussian noise ($\sigma_{\text{noise}}=0.2$ in standardized space), and retain the solution with the lowest $\mathcal{L}_{\text{inv}}$; no ground-truth parameters are used at any point.

\subsection{Cycle Consistency}
\label{sec:cycle}

The cycle-consistency constraint links the forward and inverse models by requiring that parameters recovered from a target field, when fed back through the forward model, reproduce that target field. Formally, for a ground-truth field $\mathbf{y}^*$ with mask $m$:
\begin{equation}
  \mathcal{L}_{\text{cycle}}
    = \bigl\|
        F_\theta\!\bigl(G^\star(\mathbf{y}^*),\,m\bigr)
        - \mathbf{y}^*
      \bigr\|,
  \label{eq:cycle}
\end{equation}
where $G^\star(\mathbf{y}^*)$ denotes the inverse-recovered parameters (Eq.~\ref{eq:inverse}) and $\|\cdot\|$ is a masked norm over valid mesh cells.

This constraint serves several purposes. First, it regularizes the inverse mapping: even when the forward model's parameter-to-field mapping is locally flat (so that many parameter vectors produce similar fields), the cycle loss penalizes inverse solutions whose forward reconstructions deviate from the target. Second, it enforces manifold consistency~\cite{Kustowski2020, Anirudh2020}: the
reconstructed fields $\hat{\mathbf{y}}_{\text{cycle}}=F_\theta(\hat{\mathbf{c}},m)$ are guaranteed to lie on the learned physics manifold of the forward model, ensuring physically plausible outputs. Third, the cycle loss provides a self-supervised quality metric for the inverse procedure that does not require ground-truth parameter labels at test time.

\subsection{Cycle Consistency as a Training Regularizer}
\label{sec:macc}

The optimization-based inverse of Section~\ref{sec:inverse} acts only at inference time and therefore leaves the forward model unchanged. To test whether cyclical consistency can additionally \emph{improve} the forward surrogate itself, as reported for inertial-confinement-fusion surrogates~\cite{Anirudh2020}, we introduce a learned pseudo-inverse $G_\psi:\mathbf{y}\mapsto\mathbf{c}$ and train it \emph{jointly} with the forward model under
\begin{equation}
\begin{split}
  \min_{\theta,\psi}\;
    &\underbrace{\rho\!\bigl(F_\theta(\mathbf{c},m),\mathbf{y}\bigr)}_{\text{forward}}
  + \underbrace{\rho\!\bigl(G_\psi(\mathbf{y},m),\mathbf{c}\bigr)}_{\text{inverse}}\\
  &+ \lambda_{\mathrm{cyc}}\,
    \underbrace{\bigl\|\mathbf{c}-G_\psi\!\bigl(F_\theta(\mathbf{c},m),m\bigr)\bigr\|}_{\text{cycle}},
\end{split}
  \label{eq:macc}
\end{equation}
where $\rho$ is a masked smooth-$L_1$ discrepancy. Unlike Eq.~\ref{eq:cycle}, the cycle term here back-propagates into the forward parameters $\theta$, so consistency can reshape the surrogate. The weight $\lambda_{\mathrm{cyc}}$ multiplies only the cycle term; the forward and inverse reconstruction terms are kept at unit weight, so $\lambda_{\mathrm{cyc}}=0$ recovers the forward-only baseline.

The pseudo-inverse $G_\psi$ is a compact convolutional encoder that mirrors the forward encoder in reverse. Its input is the four-channel plasma state stacked with the geometry mask as a fifth channel. Three convolutional stages increase the channel width from $24$ to $48$ to $96$; each stage applies two $3\times3$ convolutions, each followed by group normalization and a SiLU activation, and then a $2\times2$ max-pooling that halves the spatial resolution. A global average pooling collapses the final feature map to a $96$-dimensional vector, which a two-layer perceptron (hidden width $128$) maps to the five control parameters. The network exists only to seed the inference-time optimization (Sec.~\ref{sec:inverse}) and to supply the cycle-consistency signal during training; it is not used as a standalone inverse.

We quantify self-consistency with an \emph{average cyclical $R^2$} score analogous to the metric of Anirudh et al.~\cite{Anirudh2020}: for each control parameter, we sweep it linearly across its range (holding the others fixed), push the resulting parameter vectors through $F_\theta$ and then $G_\psi$, and measure the coefficient of determination between the swept and recovered values; the score is averaged over the five parameters. A value near unity indicates that the forward and inverse mappings are mutually consistent on held-out single-parameter scans.

\subsection{Loss Functions and Training}
\label{sec:training}

The forward model $F_\theta$ is trained on the SOLPS-ITER database using a composite loss function that balances pixel-level accuracy, gradient fidelity, and boundary emphasis:
\begin{equation}
  \mathcal{L} = \mathcal{L}_{\text{base}} + \lambda_w \mathcal{L}_{\text{edge}} + \lambda_g \mathcal{L}_{\text{grad}}.
  \label{eq:total_loss}
\end{equation}

The base loss $\mathcal{L}_{\text{base}}$ is a masked Huber (smooth-$L_1$) loss with $\beta = 0.05$, computed only over valid mesh cells indicated by the binary mask; per-channel weights $w_c$ allow prioritization of specific output fields, and we use $w_c = 1.0$ for $T_e, T_i$, $w_c = 1.2$ for $n_e$, and $w_c = 1.5$ for $u_a$. The edge loss $\mathcal{L}_{\text{edge}}$ is the same Huber loss reweighted by a boundary proximity map $w(\mathbf{r})$, computed as a Gaussian-decayed distance from the mask boundary ($\sigma = 3$ pixels); this emphasizes accuracy near the separatrix and target plates where gradients are steepest. Finally, $\mathcal{L}_{\text{grad}}$ is a Sobel-filter-based spatial gradient penalty that encourages the model to reproduce sharp spatial features (recycling fronts, temperature pedestals). The gradient is computed on both predicted and target fields, and the loss is the masked $L_1$ difference of the resulting gradient maps. We set $\lambda_g = 0.2$ and apply a linear warmup from epoch 20 to 60 to stabilize early training.

Training uses the Adam optimizer~\cite{adam} with initial learning rate $3 \times 10^{-4}$, \emph{ReduceLROnPlateau} scheduling (factor $0.5$, patience 5 epochs), mixed-precision (AMP) on GPU, gradient clipping at norm $1.0$, and early stopping with patience of 80 epochs. The model is trained for up to $450$ epochs with a batch size of $8$, and the best checkpoint (by validation loss) is retained. All input and output fields are standardized using training-set statistics.

\section{Results}
\label{sec:results}

Before turning to the results, we fix the metrics used throughout.
\label{sec:results_metrics}%
We quantify the \emph{accuracy} of a predicted field with four complementary measures, all computed over the valid mesh cells of the held-out test set. Writing $y_i$ for the SOLPS-ITER value and $\hat{y}_i$ for the prediction at cell $i$ (over $N$ valid cells), the mean absolute error is $\mathrm{MAE}=N^{-1}\sum_i|\hat{y}_i-y_i|$, and the normalized root-mean-square error is $\mathrm{NRMSE}=100\times\mathrm{RMSE}/(y_{\max}-y_{\min})$, the RMS error written as a percentage of the field's range. We complement these magnitude errors with two association measures: the Pearson correlation $r$ (linear agreement) and the Spearman correlation $\rho$. For the ablation of Sec.~\ref{sec:results_ablation} and the one-dimensional ensemble of Sec.~\ref{sec:profile_ensemble} we also report the coefficient of determination 
\begin{eqnarray}
    R^2=1-\sum_i(\hat{y}_i-y_i)^2/\sum_i(y_i-\bar{y})^2,
\end{eqnarray}
where $\bar{y}$ is the mean of the SOLPS-ITER values. Separately, we use \emph{robustness} to refer to the reliability of the pipeline rather than its pointwise error. At the database level it is the completion rate, the fraction of SOLPS-ITER runs that reach steady state (Sec.~\ref{sec:database}); at the model level it is the sensitivity of the prediction to input perturbations, $\mathbb{E}\,\|F_\theta(\mathbf{c})-F_\theta(\mathbf{c}+\sigma\bm{\epsilon})\|^2$ in normalized output space (Table~\ref{tab:macc_ablation}), for which a smaller value indicates a smoother, more robust surrogate.

\subsection{Forward-model accuracy}
\label{sec:results_plasma}

\begin{figure*} 
  \centering
\includegraphics[width=0.95\textwidth,height=0.85\textheight,keepaspectratio]{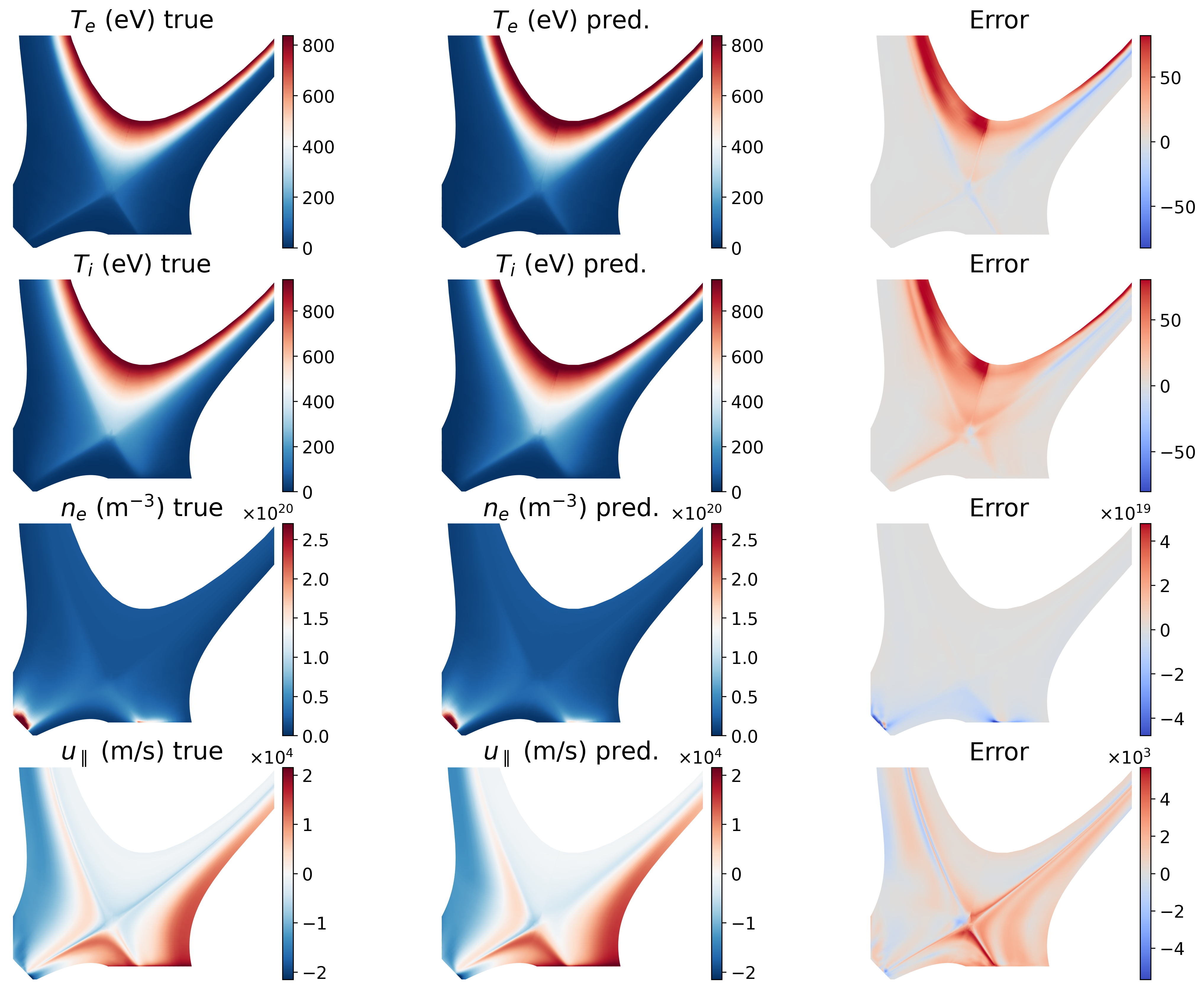}
  \caption{Comparison of SOLPS-ITER ground truth (left columns), U-Net predictions (middle columns), and absolute errors (right columns) for the four plasma-state fields on a held-out test case.Top to bottom: electron temperature $T_e$ (eV), ion temperature $T_i$ (eV), electron density $n_e$ (m$^{-3}$), and parallel velocity $u_a$ (m/s). The model captures the large-scale structure and gradients across the SOL and divertor.}
  \label{fig:plasma_fields}
\end{figure*}

Figure~\ref{fig:plasma_fields} compares SOLPS-ITER ground truth with U-Net predictions for the four plasma-state fields ($T_e$, $T_i$, $n_e$, $u_a$) in a held-out test case. The model accurately reproduces the two-dimensional structure of electron and ion temperatures, including the steep gradients near the separatrix and the characteristic decay into the scrape-off layer. Electron density is well captured across the full domain, with the core--SOL contrast and divertor compression faithfully represented. The parallel velocity field $u_a$ shows correct flow patterns from the outer midplane toward the divertor targets.

Quantitatively, the forward model achieves a global Pearson correlation exceeding $0.95$ for all four plasma fields: $r = 0.994$ for $T_e$, $r = 0.992$ for $T_i$, $r = 0.987$ for $n_e$, and $r = 0.975$ for $u_a$ (Table~\ref{tab:metrics}). Mean absolute errors are $12.2$~eV for $T_e$ and $16.7$~eV for $T_i$, which represent small fractions of the dynamic range of these fields across the database. The density field achieves an MAE of $2.6 \times 10^{18}$~m$^{-3}$ for $n_e$, while the parallel velocity MAE is $941$~m/s. Residual errors are predominantly localized near the separatrix and in the private flux region where spatial gradients are steepest.

Table~\ref{tab:metrics} summarizes the forward-model performance across all four output channels on the held-out test set.

\IfFileExists{table_metrics_paper_1.tex}{%
  \input{table_metrics_paper_1.tex}}{%
  \newcommand{\tabIImetricsbody}{%
    \rowcolor{blue!5}
    $T_e$ & eV & 12.24 & 1.26 & 0.994 & 0.996 \\
    $T_i$ & eV & 16.71 & 1.54 & 0.992 & 0.987 \\
    \rowcolor{blue!5}
    $n_e$ & m$^{-3}$ & $2.61 \times 10^{18}$ & 0.62 & 0.987 & 0.964 \\
    $u_a$ & m$\cdot$s$^{-1}$ & 940.9 & 2.54 & 0.975 & 0.976 \\
  }%
}%
\begin{table*}[!t]
\centering
\caption{Forward surrogate performance on the held-out test set (152 runs). MAE is reported in physical units, and NRMSE (\%) is the root-mean-square error normalized by the range of each field ($y_{\max}-y_{\min}$) over the valid test cells. Pearson $r$ and Spearman $\rho$ are global correlations computed over all valid points across all test samples. All metrics are defined at the start of Sec.~\ref{sec:results_metrics}.}
\label{tab:metrics}
\begin{tabular}{l l r r c c}
\toprule
\textbf{Field} & \textbf{Unit} & \textbf{MAE} & \textbf{NRMSE (\%)} & \textbf{Pearson $r$} & \textbf{Spearman $\rho$} \\
\midrule
\tabIImetricsbody
\bottomrule
\end{tabular}
\end{table*}

The four plasma state fields ($T_e$, $T_i$, $n_e$, $u_a$) all achieve Pearson correlations above $0.95$ and NRMSE below $2.6\%$, with per-sample mean Pearson values exceeding $0.98$. The parallel velocity $u_a$ is the hardest of the four to reproduce ($r=0.975$, NRMSE $2.54\%$): its sign changes along the flow, and it carries sharper spatial structure than the temperatures and density, so residual errors concentrate near the separatrix and the divertor targets.
\subsection{Model size and inference speed}
\label{sec:results_complexity}

The conditional U-Net has ${\sim}4.3 \times 10^6$ trainable parameters with a base filter width of 48, occupying $16.5$~MB in single precision. Each forward pass maps the single-channel mask input, with the five scalar parameters injected via FiLM, to four output channels at $36 \times 96$ resolution. On a single CPU core, inference takes $16$~ms per sample; on a GPU, sub-millisecond throughput is achievable in batched mode. With GPU inference, these latencies are compatible with real-time plasma control loops ($1$--$10$~ms cycle times)~\cite{Degrave2022,Seo2024,StJohn2021}, digital-twin frameworks~\cite{Tang2024,Willcox2024}, and Monte Carlo uncertainty quantification over the input space.

\subsection{Invertibility gains from cycle consistency}
\label{sec:results_ablation}

\begin{figure*} 
  \centering
  \includegraphics[width=\linewidth]{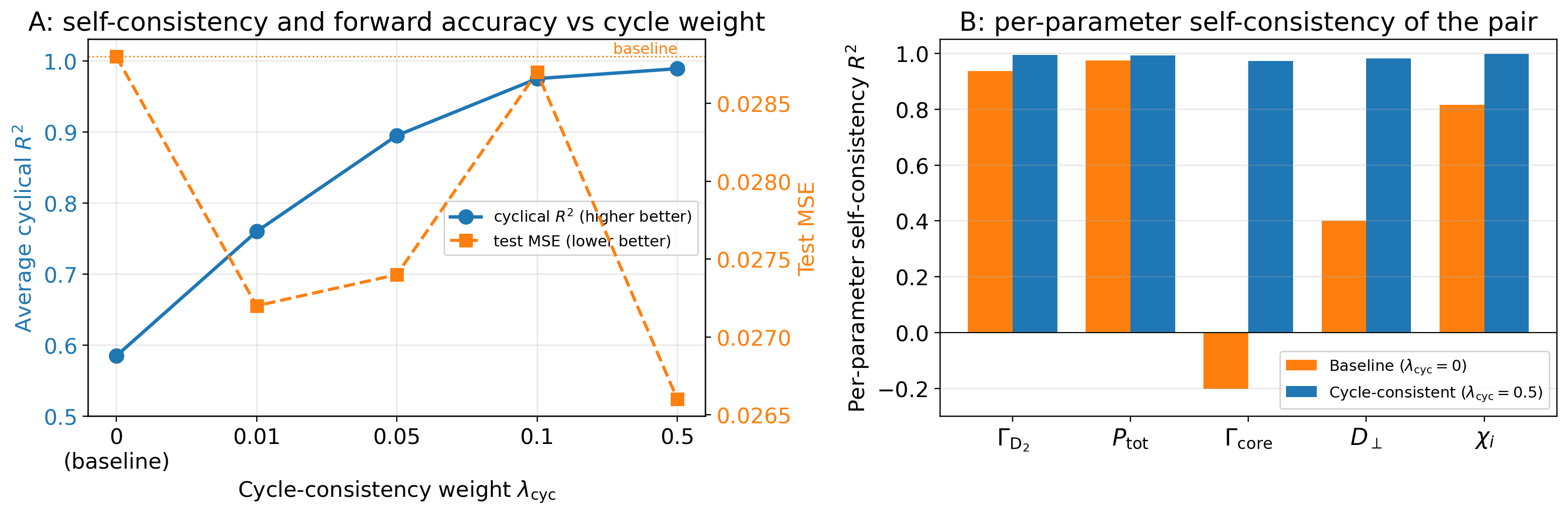}
  \caption{Cycle-consistency ablation. (A)~Average cyclical $R^2$ (left axis, higher is better) and held-out test MSE (right axis, lower is better) as the cycle weight $\lambda_{\mathrm{cyc}}$ is increased. Self-consistency increases from $0.59$ at the forward-only baseline ($\lambda_{\mathrm{cyc}}=0$) to $0.99$. At the same time, the test MSE stays at or below the baseline (dotted line) throughout, so self-consistency improves at no cost to forward accuracy. (B)~Per-parameter self-consistency of the trained forward/inverse pair for the forward-only baseline ($\lambda_{\mathrm{cyc}}=0$, orange) versus the cycle-consistent model ($\lambda_{\mathrm{cyc}}=0.5$, blue). The cycle term brings every control parameter above $0.97$, including $\Gamma_{\mathrm{core}}$ and $D_\perp$, which are otherwise weakly recovered by the pair. Higher $R^2$ is better; lower MSE is better. Each $\lambda_{\mathrm{cyc}}$ is trained from scratch, so the intermediate points carry run-to-run variation; the baseline ($\lambda_{\mathrm{cyc}}=0$) and the adopted $\lambda_{\mathrm{cyc}}=0.5$ are the comparison of interest. Exact sweep values are given in Table~\ref{tab:macc_ablation}.}
  \label{fig:macc_ablation}
\end{figure*}

To test whether cyclical consistency improves the forward surrogate, and not merely the inverse, we sweep the regularization weight
$\lambda_{\mathrm{cyc}}$ in the joint objective (Eq.~\ref{eq:macc}) and, for each value, retrain the forward model with the learned pseudo-inverse from scratch. Figure~\ref{fig:macc_ablation}(A) and Table~\ref{tab:macc_ablation} report the held-out test mean squared error and the average cyclical $R^2$ score as $\lambda_{\mathrm{cyc}}$ is increased, following the ablation protocol of Anirudh et al.~\cite{Anirudh2020}. As $\lambda_{\mathrm{cyc}}$ increases, the forward and inverse networks become markedly more cyclically self-consistent: the average cyclical $R^2$ increases from $0.59$ at $\lambda_{\mathrm{cyc}}=0$ to $0.99$ at $\lambda_{\mathrm{cyc}}=0.5$. Crucially, this improved self-consistency does not come at the expense of forward accuracy: the held-out test MSE remains comparable to or below the forward-only baseline ($\lambda_{\mathrm{cyc}}=0$) across the entire sweep. The prediction's sensitivity to input perturbations (Table~\ref{tab:macc_ablation}) remains bounded and comparable to the unregularized baseline across the range explored.

The effect is clearest at the level of individual control parameters. Figure~\ref{fig:macc_ablation}(B) compares the per-parameter recovery $R^2$ of the forward-only baseline with that of the consistency-regularized model ($\lambda_{\mathrm{cyc}}=0.5$). Without the cycle term, the trained pair is self-consistent for some control parameters but not others: the recovered $R^2$ for the core particle source $\Gamma_{\mathrm{core}}$ and the cross-field diffusivity $D_\perp$ is low, whereas the input power $P_{\mathrm{tot}}$ and the gas-puff rate $\Gamma_{\mathrm{D}_2}$ are already well recovered. Training with the cycle term makes the pair self-consistent across all five parameters, each above $0.97$. We emphasize that this improved recovery is a property of the jointly trained forward/inverse \emph{pair} rather than of the forward model in isolation: it does not by itself imply that the forward model has become more invertible. We use it only as a self-supervised consistency signal and adopt $\lambda_{\mathrm{cyc}}=0.5$, which is both the most self-consistent and the most accurate on held-out data.

\begin{table}[!b]
\centering
\caption{Cycle-consistency ablation sweep (760-run database; 608 training / 152 test runs, trained from scratch at each $\lambda_{\mathrm{cyc}}$). As the cycle weight increases, the average cyclical $R^2$ increases toward unity while the held-out test MSE stays comparable to or below the forward-only baseline ($\lambda_{\mathrm{cyc}}=0$). Sensitivity is the mean squared change of the normalized prediction under Gaussian input perturbations ($\sigma=0.1$); a smaller value indicates a more robust surrogate. Best values in bold.}
\label{tab:macc_ablation}
\setlength{\tabcolsep}{4pt}
\resizebox{\columnwidth}{!}{%
\begin{tabular}{@{}l c c c@{}}
\toprule
$\lambda_{\mathrm{cyc}}$ & Test MSE & Cyclical $R^2$ & Sens. ($\times10^{-3}$) \\
\midrule
\rowcolor{blue!5}
$0$ (baseline) & $0.0288$ & $0.585$ & $\mathbf{2.14}$ \\
$0.01$ & $0.0272$ & $0.760$ & $3.93$ \\
\rowcolor{blue!5}
$0.05$ & $0.0274$ & $0.895$ & $3.97$ \\
$0.1$  & $0.0287$ & $0.975$ & $5.10$ \\
\rowcolor{blue!5}
$0.5$  & $\mathbf{0.0266}$ & $\mathbf{0.989}$ & $3.51$ \\
\bottomrule
\end{tabular}}
\end{table}

\subsection{Inverse parameter recovery}
\label{sec:results_cycle}

We evaluate the inverse model on held-out test cases using the optimization-based inference procedure (Sec.~\ref{sec:inverse}). To initialize the optimization we use the learned pseudo-inverse $G_\psi$ of Sec.~\ref{sec:macc}: its convolutional encoder maps the observed fields, with the geometry mask supplied as an auxiliary channel, to a parameter estimate $\hat{\mathbf{c}}_0=G_\psi(\mathbf{y}^*,m)$ that provides the warm start; no ground-truth parameters are used. Starting from this estimate, the optimizer refines the parameters by minimizing the forward-model discrepancy (Eq.~\ref{eq:inverse}) through the frozen $F_\theta$, with $N_r=5$ random restarts per case and cosine learning-rate annealing over 1200 Adam steps. The cycle-consistency metric $\mathcal{L}_{\text{cycle}}$ (Eq.~\ref{eq:cycle}) is then evaluated by passing the recovered parameters through $F_\theta$ and comparing the reconstructed fields against the original targets.

This procedure provides two complementary diagnostics: (i) the accuracy of parameter recovery (how close $\hat{\mathbf{c}}$ is to the true  $\mathbf{c}^*$), and (ii)~the cycle reconstruction quality (how well  $F_\theta(\hat{\mathbf{c}},m)$ matches~$\mathbf{y}^*$). These need not agree: the inverse problem in edge-plasma modeling can be ill-posed, since the forward mapping is locally insensitive to certain parameter combinations, so good cycle reconstruction does not on its own guarantee accurate recovery. Reporting both diagnostics lets us separate the two and determine whether a clean field reconstruction reflects a genuinely well-identified parameter set.

\begin{figure*} 
    \centering
    \includegraphics[width=\linewidth]{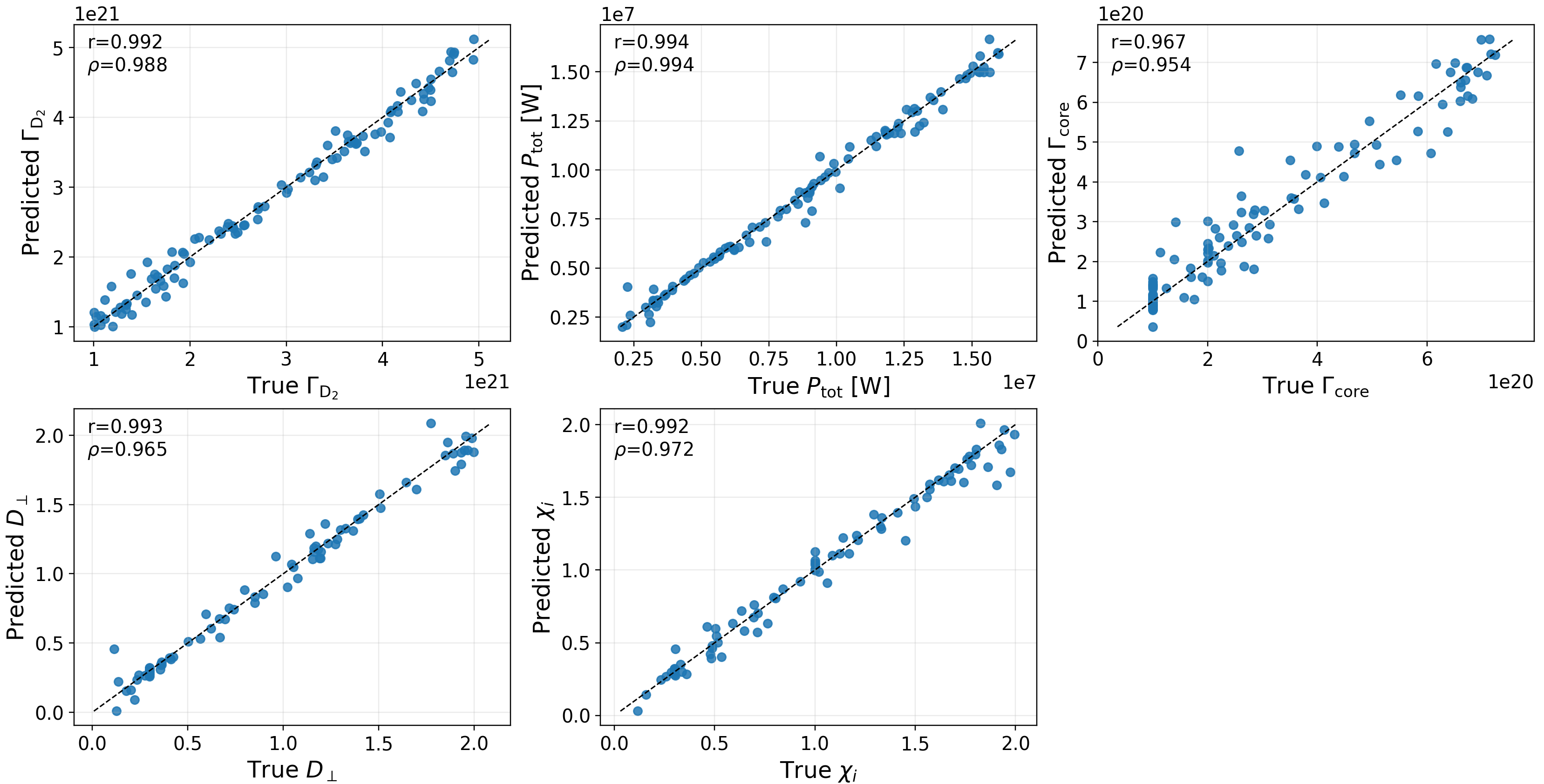}
    \caption{Inverse parameter recovery on held-out test cases. Each panel shows the true versus recovered value for one control parameter, obtained by optimizing through the frozen forward model with multiple random restarts. Pearson and Spearman correlation coefficients are annotated per parameter.}
    \label{fig:inverse_param}
\end{figure*}

Figure~\ref{fig:inverse_param} shows the true versus recovered values for the five control parameters on held-out test cases. All five are recovered well, with Pearson correlations at or above $0.97$: $P_{\mathrm{tot}}$ ($r=0.99$, $\rho=0.99$), $\Gamma_{\mathrm{D}_2}$ ($r=0.99$, $\rho=0.99$), $D_\perp$ ($r=0.99$, $\rho=0.97$), $\chi_i$ ($r=0.99$, $\rho=0.97$), and the core fueling $\Gamma_{\mathrm{core}}$ ($r=0.97$, $\rho=0.95$). The core fueling remains the most scattered of the five, consistent with the weaker sensitivity of the downstream SOL and divertor fields to it. Still, it is well identified: the inverse optimization works through the frozen forward model. Hence, a parameter to which that model is only weakly sensitive is the hardest to pin down. Reporting both parameter-recovery accuracy and cycle-reconstruction quality is what distinguishes a genuinely well-identified parameter from a degenerate one.

\section{Uncertainty-Quantified Profile Ensemble for Control and Active Learning}
\label{sec:profile_ensemble}

Alongside the two-dimensional U-Net, the framework's second surrogate is a standalone query-by-committee ensemble of multi-task regression networks that predicts one-dimensional profiles of electron temperature $T_e$ and density $n_e$ at the outboard midplane and both divertor targets directly from the five control parameters, as a function of the normalized poloidal flux $\psi_N$. We restrict the targets to $T_e$ and $n_e$ deliberately: these are the independent state variables measured directly by Thomson scattering and are the quantities most relevant to edge diagnostics and real-time control. Derived quantities such as the divertor heat flux are not given a separate learned head, since they are constitutive functions of $T_e$ and $n_e$ and predicting them independently would risk thermodynamic inconsistency; when needed, they can instead be evaluated from the predicted profiles with uncertainty propagated through the ensemble. Where the U-Net provides physically complete fields for design and inverse analysis, this ensemble targets a different deployment mode: fast, pointwise profile evaluation with \emph{uncertainty estimates}, as required for real-time control and for active-learning selection of the most informative new SOLPS-ITER runs~\cite{Dasbach2023,Zhu2022}. 

SOLPS-ITER simulations are large (many cells, many time steps), so we favor \emph{parametric} models whose evaluation cost does not grow with the size of the training set. We also prefer methods that admit a simple, effective UQ scheme. We therefore use an ensemble of neural networks and take the ensemble variance as a proxy for model uncertainty. This is a practical example of the query-by-committee (QBC) approach~\cite{QBC}, an active-learning strategy that proposes new training points where a committee of models disagrees the most. We have used this approach in previous work~\cite{DiawPRE20,DiawNME2022}.

Training minimizes mean-squared error loss,
\[
\mathcal{L} \;=\; \sum_{i,t}\,\big(\hat{y}^{(t)}_{i}-y^{(t)}_{i}\big)^2
\]
using Adam~\cite{adam} in mini-batches of size 256, with early stopping (patience 20 epochs, learning rate halving) and a cap of 400 epochs. We form the ensemble by bootstrapping: each member is trained on a different random split, with $10\%$ held out for early-stopping validation and a further held-out fold used to score it. Members that do not reach $R^2\ge 0.90$ on every output channel of that fold are discarded. We retain $n_{\rm ensemble}=5$ trained models. Because $n_e$ spans several decades and the target $T_e$ ranges from below $1$ to above $200$~eV, both are trained on their $\log_{10}$ values and mapped back to physical units for evaluation.

To assess the trustworthiness of the model prediction, we threshold the committee disagreement with a quality score
\begin{equation}
    s_i \;=\; k\,\frac{\sigma_i}{S},
\end{equation}
where $\sigma_i$ is the ensemble standard deviation at point $i$ and $k=2$. The global scale $S = \bar{\sigma}\,\bar{E}_{\rm cal}/\bar{E}_{\rm ens}$ rescales the raw committee spread to the magnitude of the actual prediction error: $\bar{\sigma}$ is the mean ensemble standard deviation, $\bar{E}_{\rm cal}$ the mean absolute error of the ensemble-mean prediction, and $\bar{E}_{\rm ens}$ the mean ensemble spread, all computed on the held-out validation fold. If $s_i\ge 1$, the point is flagged for further simulation.

\begin{figure*} 
  \centering
  \includegraphics[width=\linewidth]{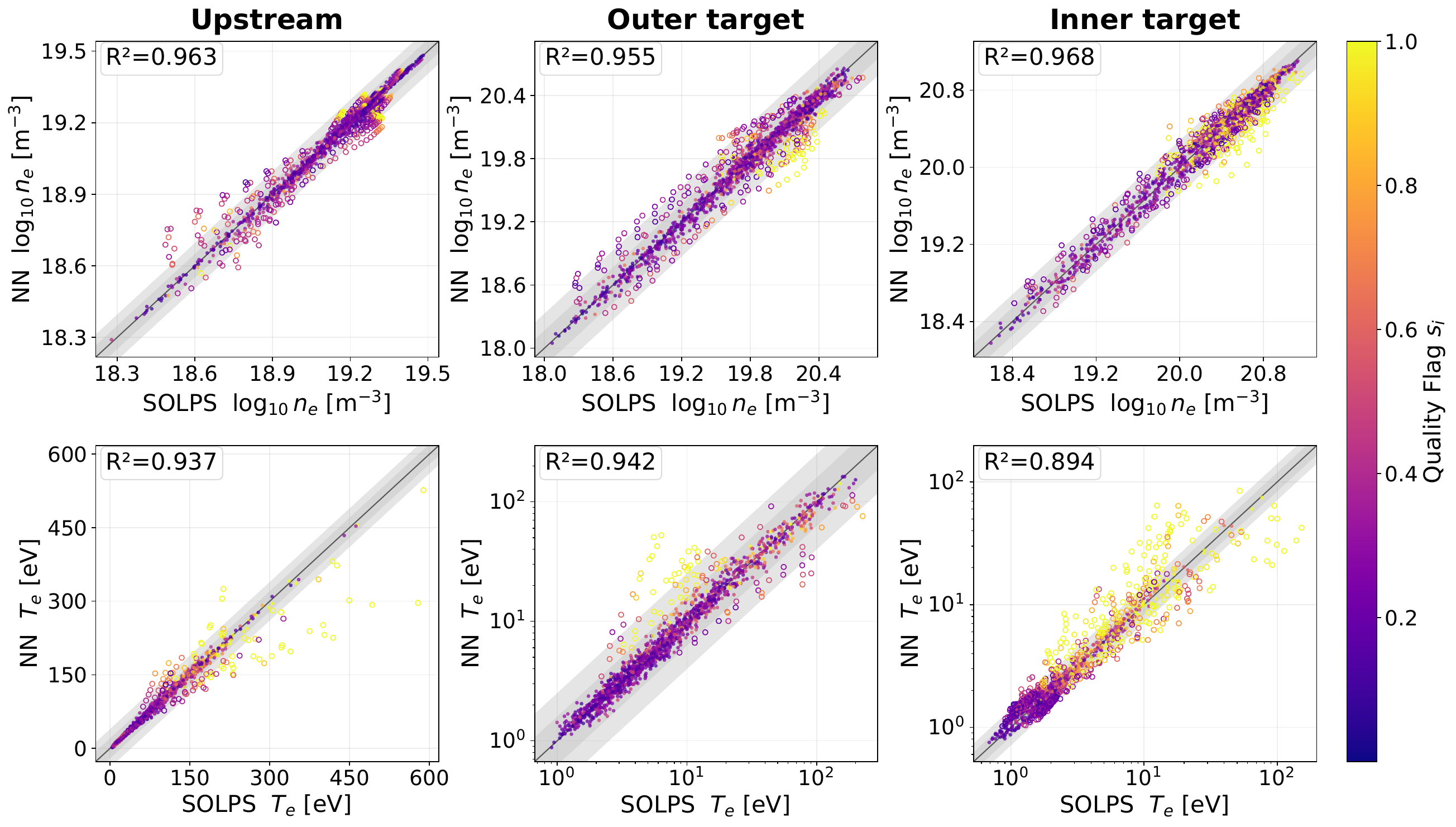}
  \caption{Evaluation of the one-dimensional profile ensemble. Columns correspond to the three profile locations (upstream/outboard midplane, outer divertor target, and inner divertor target); rows correspond to the two predicted quantities, electron density $n_e$ and electron temperature $T_e$, on the independent test set ($3040$ points per location; $12160$ training points). Target $T_e$ is shown on a logarithmic axis; the per-panel coefficient of determination is annotated in each panel. All points are colored by the quality flag $s_i$; open circles mark the highest-uncertainty points ($s_i>1$) and are flagged for additional SOLPS-ITER simulation.}
  \label{fig:NN:PERFORMANCE}
\end{figure*}

We train each ensemble on $12160$ points ($608$ SOLPS-ITER runs) and evaluate it on $3040$ held-out test points ($152$ runs), using the same $80/20$ train/test split as the 2D surrogate so that the two models are directly comparable. The performance on the held-out test set is shown in Fig.~\ref{fig:NN:PERFORMANCE}. The ensemble reproduces the held-out $T_e$ and $n_e$ profiles with high fidelity, with the per-panel coefficient of determination annotated in Fig.~\ref{fig:NN:PERFORMANCE}. Points are colored by the quality flag $s_i$; those selected for SOLPS-ITER verification ($s_i> 1$) are drawn as open circles, highlighting the regions where the network is least confident.

Among the three locations, the inner divertor target is reproduced least accurately, with the lowest $T_e$ coefficient of determination ($R^2\approx 0.89$, versus $\gtrsim 0.94$ at the upstream and outer-target locations) and by far the largest fraction of flagged points ($\sim 13\%$, against $\sim 3\%$ at the outer target). This is consistent with edge physics rather than a shortcoming of the network: the inner leg detaches at lower upstream density than the outer, so across the database it spans both attached and detached states, and near the strike point the target $T_e$ becomes effectively bimodal and steeply varying, a harder mapping to learn from the control parameters alone. The committee makes this visible rather than hiding it: the inner target carries the largest ensemble disagreement. It is therefore the top priority for active-learning acquisition of new SOLPS-ITER runs.

\section{Discussion}
\label{sec:discussion}

Our framework builds on the manifold-and-cycle-consistency paradigm for surrogate modeling of tokamak edge plasmas. Anirudh et al.~\cite{Anirudh2020} enforce manifold consistency with a Wasserstein autoencoder that learns a low-dimensional latent representation; we instead use a conditional U-Net whose skip connections preserve fine spatial detail. Both share the core insight that cycle consistency regularizes the inverse model, and in both, the reconstructed outputs are constrained to lie on the learned physics manifold. The tokamak setting adds its own difficulties: the mesh geometry is irregular, which is what forces the masking described above, and the fields carry hundreds of sharp recycling fronts that do not admit a smooth, low-dimensional structure.

Table~\ref{tab:metrics} shows that all four fields are reproduced accurately (Pearson correlation $>0.95$), with a mild hierarchy among them: the temperatures and density, which vary smoothly over the scale of the scrape-off layer (SOL) width, are captured most efficiently by the convolutional receptive field, whereas the parallel velocity $u_a$ carries sharper spatial structure and a sign change along the flow and is therefore the hardest to fit. That same spatial structure is what motivates the convolutional inductive bias: coherent large-scale flow patterns like $u_a$ are reproduced far more accurately by a model that shares information between neighboring cells than by a pointwise regressor. This supports the choice of a deterministic, parameter-conditioned convolutional operator for the data-scarce regime studied here, where generative alternatives would have to learn a full distribution from only a few hundred samples.

The millisecond-scale inference described in Sec.~\ref{sec:results_complexity} enables several practical applications: (i) rapid exploration of the five-dimensional parameter space for sensitivity analysis and scenario optimization~\cite{DiawPRE20, Noack:2019}; (ii) transport-coefficient and parameter inference from diagnostic measurements through the differentiable inverse procedure, replacing iterative manual SOLPS-ITER fitting~\cite{Wilcox2026}; and (iii) real-time or near-real-time edge-plasma state estimation for control applications~\cite{Degrave2022,Seo2024}, conditioned on a limited set of diagnostic measurements. Comparable digital-twin paradigms have been demonstrated for particle accelerator control~\cite{Roussel2023,StJohn2021}.

Generating the SOLPS-ITER database is computationally demanding: each run solves a stiff, coupled nonlinear system whose convergence trajectory is highly sensitive to the initial state. The $k$-d tree restart strategy (Sec.~\ref{sec:database}) functions as a discrete form of numerical continuation~\cite{AllgowerGeorg2003}, where the solution at one parameter-space point predicts the solution at a nearby point. Classical continuation traces a single curve through parameter space; here, the $k$-d tree generalizes this concept to a multi-dimensional Latin hypercube design~\cite{McKay1979} by selecting, for each new sample, the nearest previously converged neighbor in $(\log N)$ time~\cite{Bentley1975,Friedman1977}. The database grows as a branching process: starting from a single seed, each newly converged case extends the frontier of reliable initial conditions, progressively reducing the parameter-space distance to subsequent targets. Beyond this branching construction, the warm-start strategy improves robustness: in a prior cold-started ensemble, roughly $30\%$ of runs initiated from generic initial conditions diverged outright, and a further fraction never reached steady state~\cite{Dasbach2023}, whereas the branching approach achieved a completion rate above $95\%$ across the full five-dimensional design, yielding the $762$-run database used here. This infrastructure is essential for any surrogate relying on a large, uniformly sampled training set; without it, gaps in the database would directly translate into blind spots in the learned model.

Several limitations remain, and we state them plainly. First, the current database covers only a single DIII-D lower-single-null configuration with deuterium-only fueling and spatially uniform transport coefficients; extending it to other machines, geometries, plasma mixtures, and spatially varying transport is necessary for validation and, perhaps, for transfer learning. Second, the surrogate has not yet been validated against experimental DIII-D measurements, a critical step before deployment. Third, physics constraints beyond admissibility, such as global radiated power and $Z_{\text{eff}}$ consistency, and the conservation of particles, momentum, and energy, are not yet enforced; imposing them is a natural way to sharpen the physical fidelity of the surrogate.

Finally, although the current inverse procedure is warm-started by the learned pseudo-inverse $G_\psi$ (Sec.~\ref{sec:macc}), the refinement still requires gradient-based optimization through the frozen forward model; end-to-end joint training with cycle consistency could further improve convergence speed and inverse accuracy.

\section{Conclusions}
\label{sec:conclusions}

A cycle-consistent neural surrogate framework is introduced for modeling two-dimensional SOLPS-ITER edge-plasma fields. The framework incorporates a $k$-d tree warm-start strategy that selects nearest-neighbor restarts from a database of previously converged SOLPS-ITER simulations, thereby reducing both simulation failure rates and wall-clock time: across the five-dimensional parameter scan, the warm start reached a completion rate above $95\%$, yielding $762$ converged runs, whereas roughly $30\%$ of comparable cold-started runs have been reported to fail outright, with more never reaching steady state~\cite{Dasbach2023}.

A parameter-conditioned U-Net maps scalar control parameters directly to the two-dimensional plasma state ($T_e$, $T_i$, $n_e$, $u_a$), and achieves Pearson correlation coefficients exceeding $0.95$ for all four plasma state variables. Cycle-consistency regularization is shown to yield a self-consistent forward/inverse pair at negligible cost to forward accuracy, supplying a self-supervised consistency signal that requires no ground-truth labels. Finally, an inverse inference procedure coupled with cycle-consistency regularization enables parameter recovery from target fields. It ensures that reconstructed outputs lie on the learned physics manifold, providing a self-supervised reliability metric that requires no ground-truth labels at inference time. For held-out cases, the inverse recovers all five control parameters with Pearson correlations of $0.97$ or higher, including the cross-field transport coefficients $D_\perp$ and $\chi_i$ (both $r\approx0.99$); since these are typically set by hand-tuned SOLPS-ITER fits to diagnostics, the differentiable inverse offers an automated route toward transport-coefficient inference. A companion query-by-committee profile ensemble supplies predictive uncertainty estimates whose committee disagreement concentrates on the physically hardest regime, the near-strike-point inner divertor closest to detachment (flagging roughly $13\%$ of inner-target points against $3\%$ at the outer target), and thereby nominates the most informative new SOLPS-ITER runs for active-learning acquisition.

Looking ahead, several directions are being pursued. First, validation against experimental DIII-D measurements from discharge 174310 is underway, which will test the surrogate's ability to generalize beyond the simulation database. Second, extending the inverse procedure to operate on sparse one-dimensional diagnostic signals (e.g., Thomson scattering profiles) would directly replace the manual transport-coefficient fitting workflow of interpretive boundary modeling~\cite{Wilcox2026}. Third, incorporating additional physics constraints (global radiated power $P_{\text{rad}}$ and effective charge $Z_{\text{eff}}$) and extending to spatially varying transport coefficients will improve the surrogate's physical fidelity and applicability. Finally, extending the approach to other magnetic configurations and machines (e.g., ITER, SPARC, MAST-U) will test its generality. The companion neutral-source model, which maps the plasma state predicted here to EIRENE source terms for inline replacement of the kinetic neutral step, is developed separately.

\section*{CRediT authorship contribution statement}

{\bf A. Diaw}: Conceptualization, Methodology, Software, Formal analysis,
Investigation, Data curation, Visualization, Writing -- original draft. {\bf S. De Pascuale}: Conceptualization, Methodology.
{\bf J.-S. Park}: Conceptualization, Methodology. {\bf I. Paradela Perez}: Conceptualization, Methodology.
{\bf J.D. Lore}: Software, Resources, Methodology, Funding acquisition.
{\bf S. Dasbach}: Conceptualization, Methodology.

\section*{Code and Data availability}
The code is available at \url{https://github.com/abdoudiaw/solpex}. The sampled data is archived on figshare at \url{https://doi.org/10.6084/m9.figshare.32048490}.

\section*{Acknowledgments}

This work was supported by the U.S. Department of Energy (DOE), Office of Science, Office of Fusion Energy Sciences. This research used resources of the Oak Ridge Leadership Computing Facility at Oak Ridge National Laboratory, which is supported by DOE under Contract DE-AC05-00OR22725. This research used resources of the Oak Ridge National Laboratory Research Cloud. The results are obtained with the help of the EIRENE package (see www.eirene.de) including the related code, data and tools \cite{Reiter2005}. DIFFER is part of the institutes organisation of NWO.

\section*{Declaration of generative AI and AI-assisted technologies in the writing process}

During the preparation of this work, the authors used Claude (Anthropic) to improve the manuscript's readability and language. After using this tool, the authors reviewed and edited the content as needed and took full responsibility for the published article.

\appendix

\section{$k$-d Tree Warm-Start}
\label{sec:appendix_warmstart}

\begin{figure}[!t]
  \centering
  \includegraphics[width=\linewidth]{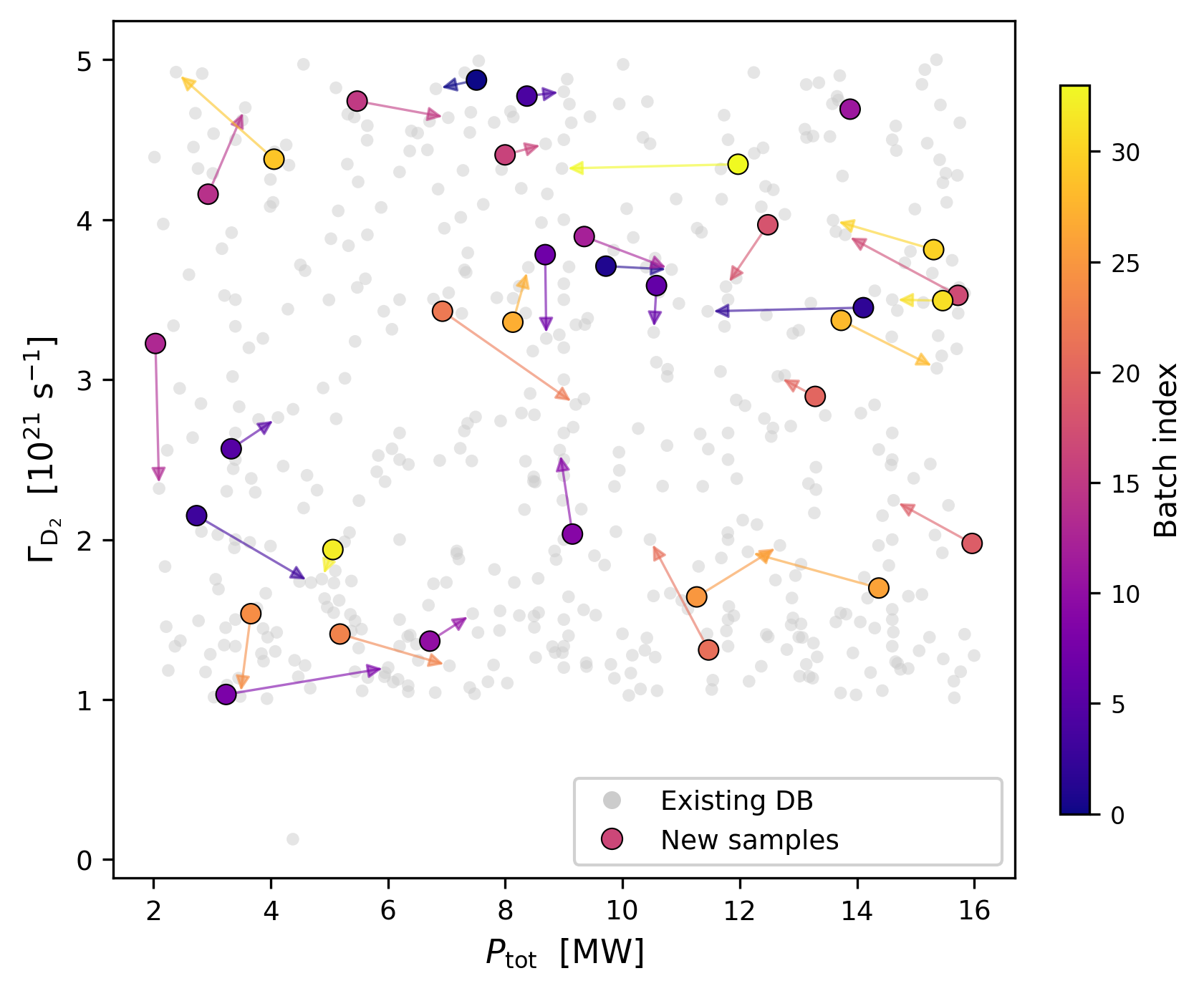}
  \caption{The $k$-d tree warm-start strategy applied during database generation, projected onto the $(P_{\mathrm{tot}},\,\Gamma_{\mathrm{D_2}})$ plane. Grey points are the existing converged database; colored markers are new samples, each initialized from its $k$-d tree nearest neighbor among the already-converged runs. Arrows link each new sample to the neighbor it was warm-started from, and marker color encodes the batch index, so the branching growth of the database frontier is visible as sampling progresses.}
  \label{fig:kdtree}
\end{figure}

\textbf{Construction.} We build a balanced $k$-d tree over all LHS design points $\{\mathbf{x}_i\}$ by recursively splitting the data along coordinate directions. This structure supports nearest-neighbor queries in $\mathcal{O}(\log N)$ time on average for $d \ll N$, where $d=5$ is the number of input parameters and $N$ is the number of design points. Given a target input $\mathbf{t}_\star$, we define a weighted Euclidean distance
\begin{equation}
  d_w^2(\mathbf{x}_i,\mathbf{t}_\star)
    = \sum_{j=1}^{5} w_j \left( x_{i,j} - t_{\star,j} \right)^2,
  \label{eq:kdtree-distance}
\end{equation}
where $w_j$ is a user-defined weight for the $j$-th parameter (here $w_j = 1$ for all $j$). For each target, we query the tree for the nearest already-converged neighbor and initialize the SOLPS-ITER state from that solution rather than from a generic seed (Fig.~\ref{fig:kdtree}). The database thus grows as a branching process. Each newly converged case extends the frontier of reliable initial conditions, a discrete form of numerical continuation~\cite{AllgowerGeorg2003} generalized to a multi-dimensional Latin hypercube design~\cite{McKay1979}.

\section{Database Convergence Detection}
\label{sec:appendix_convergence}

Determining whether a SOLPS-ITER run has reached steady state is nontrivial. Density, momentum, energy, and radiation fields relax toward equilibrium at different rates, and the iterative coupling between B2.5 and EIRENE introduces correlated fluctuations that persist over many timesteps. We address this using autocorrelation and block-averaging techniques from equilibrium sampling in molecular dynamics~\cite{AllenTildesley2017, FrenkelSmit2002}. The same machinery underpins Green-Kubo transport-coefficient calculations in molecular dynamics simulations \cite{Diaw15,Ticknor2016}; here, we repurpose it to estimate the correlation time of SOLPS-ITER diagnostics and construct a statistically grounded convergence criterion~\cite{FlyvbjergPetersen1989,Sokal1997}.

We monitor four scalar diagnostics extracted at each SOLPS-ITER timestep: the electron and ion temperatures at the outboard-midplane separatrix ($T_{e,\mathrm{sep}}^{\mathrm{OMP}}$, $T_{i,\mathrm{sep}}^{\mathrm{OMP}}$), the electron density at the same location ($n_{e,\mathrm{sep}}^{\mathrm{OMP}}$), and the total number of particles ($\langle tmne \rangle$). Each diagnostic is treated as a noisy relaxation process whose fluctuations encode the residual coupling between the plasma and neutral subsystems.

Given a time series $\{q_k\}_{k=0}^{N-1}$ with sample mean $\bar{q}$, we compute the normalized autocorrelation function
\begin{equation}
  C(\ell)
  = \frac{\displaystyle
      \sum_{k=0}^{N-\ell-1} (q_k - \bar{q})\,(q_{k+\ell} - \bar{q})
    }{\displaystyle
      \sum_{k=0}^{N-1} (q_k - \bar{q})^2
    },
  \qquad \ell = 0,1,2,\dots,
  \label{eq:autocorr}
\end{equation}
so that $C(0) = 1$ and $|C(\ell)| \le 1$ for $\ell>0$. The integral correlation time is then estimated as
\begin{equation}
  \tau_{\mathrm{int}} \approx \Delta t
  \left( \frac{1}{2} + \sum_{\ell=1}^{\ell_{\max}} C(\ell) \right),
  \label{eq:tau-int}
\end{equation}
where the sum is truncated at the first non-positive value of $C(\ell)$ or at a prescribed maximum lag~\cite{Sokal1997}. This quantity plays the same role as the correlation time in MD block averaging~\cite{FlyvbjergPetersen1989}: it sets the minimum window length over which successive block means can be treated as approximately independent.

\begin{figure*}[!t]
    \centering
    \includegraphics[width=\linewidth]{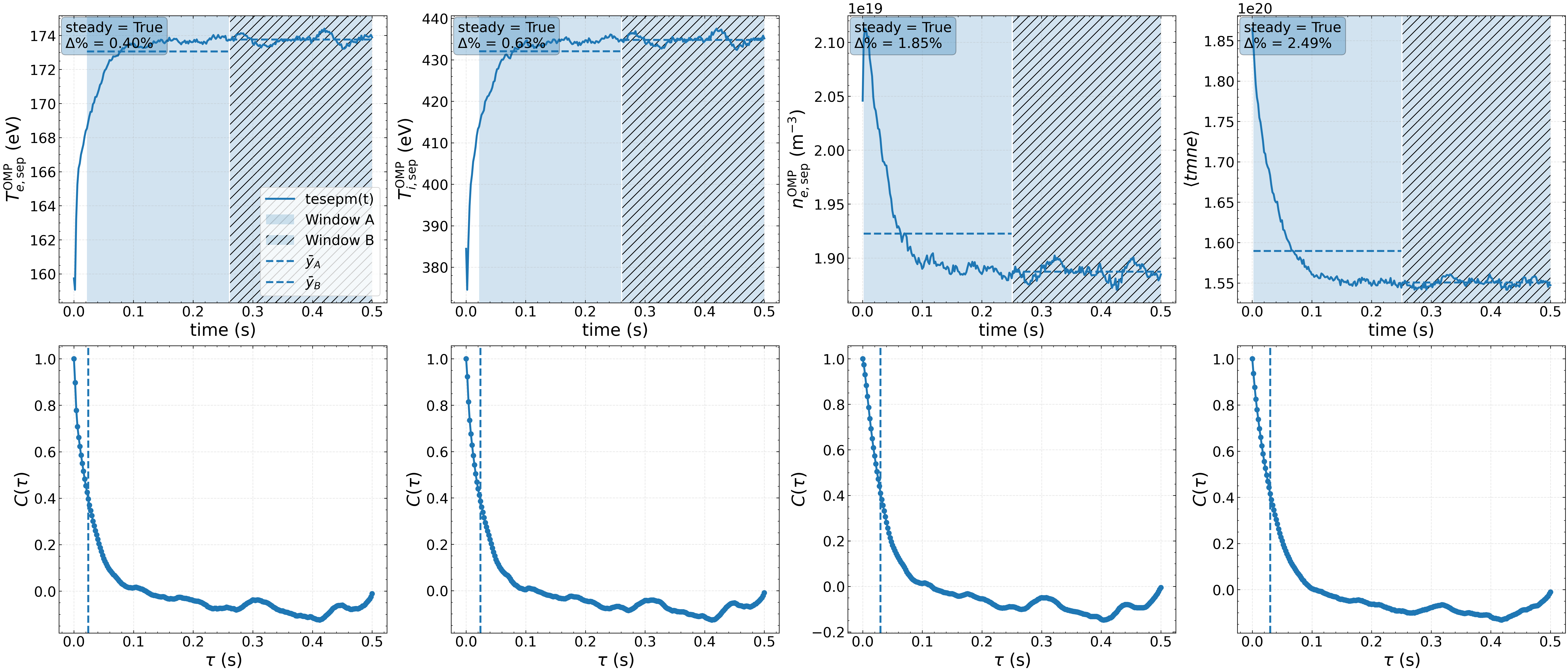}
    \caption{Steady-state diagnostics for a representative SOLPS-ITER run. Top row: time series of the four monitored quantities with windows A and B used for the drift test. Bottom row: normalized autocorrelation functions with the estimated integral correlation time $\tau_{\mathrm{int}}$ (dashed line).}
    \label{fig:convergence1}
\end{figure*}

Using $\tau_{\mathrm{int}}$, we define a tail region consisting of the last $T_{\mathrm{tail}} = N_{\mathrm{tail}}\,\tau_{\mathrm{int}}$ timesteps of the series and split it into two non-overlapping windows of equal length $L = N_{\mathrm{win}}\,\tau_{\mathrm{int}}$, denoted A and B. The drift between the two window means, $\Delta \bar{q} = \bar{q}_B - \bar{q}_A$, is compared against its expected statistical fluctuation,
\begin{equation}
  \sigma_{\Delta \bar{q}} \approx
  \sqrt{2}\,\sqrt{\frac{2\,\tau_{\mathrm{int}}}{L}}\;\sigma_q,
\end{equation}
where $\sigma_q$ is the standard deviation within each window. A quantity is declared to be in a steady state when
\begin{equation}
  \frac{|\Delta \bar{q}|}{|\bar{q}_B|}
  < \varepsilon_{\mathrm{rel}}
  \quad\text{and}\quad
  \frac{|\Delta \bar{q}|}{\sigma_{\Delta \bar{q}}}
  < z_{\max},
  \label{eq:steady-criterion}
\end{equation}
with $\varepsilon_{\mathrm{rel}} = 0.01$ and $z_{\max}=2$. The first condition ensures that the mean has not drifted by more than $1\%$; the second ensures that the observed drift is statistically consistent with equilibrium fluctuations at the estimated correlation time. A simulation is accepted as converged only when all four diagnostics simultaneously satisfy Eq.~\eqref{eq:steady-criterion}. Figure~\ref{fig:convergence1} illustrates the procedure for a representative run, showing the time series, the two comparison windows, and the corresponding autocorrelation functions. This approach gives a consistent steady-state check inline while SOLPS-ITER is running. 

\section{One-Dimensional Profile Cycle Consistency}
\label{sec:appendix_profile_cycle}

\begin{figure*}[!t]
  \centering
  \includegraphics[width=\linewidth]{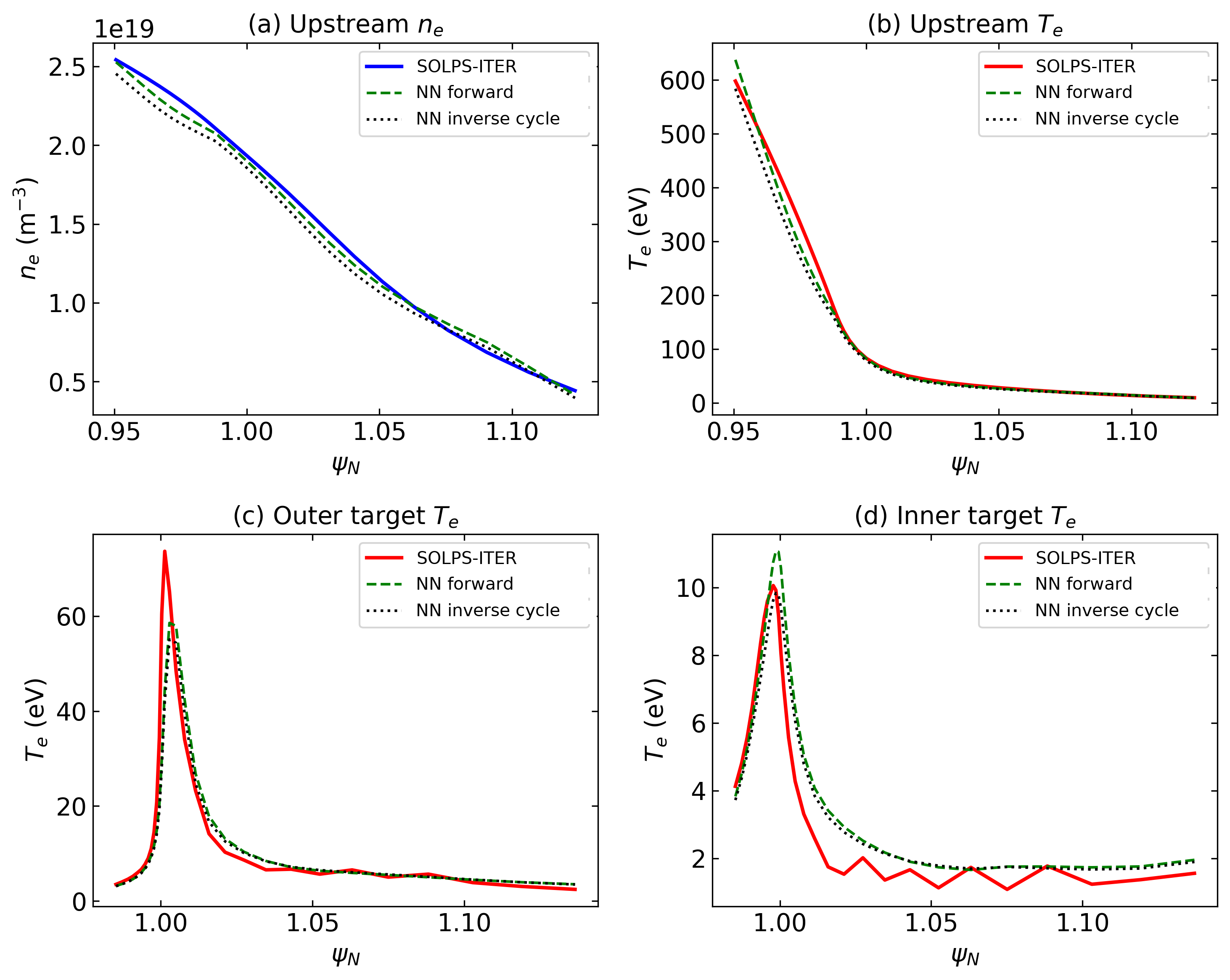}
  \caption{Predicted electron profiles as a function of normalized poloidal flux $\psi_N$ for a representative held-out test case. \emph{Solid lines}: SOLPS-ITER ground truth. \emph{Dashed lines} (``NN forward''): the forward surrogate $F_\theta$ evaluated at the \emph{true} control parameters, isolating forward-model accuracy. \emph{Dotted lines} (``NN inverse cycle''): the cycle reconstruction, in which the control parameters are first recovered from the SOLPS-ITER fields by the inverse procedure $G^\star$ and then propagated back through the forward model. Panels show (a) upstream (outboard-midplane) $n_e$, (b) upstream $T_e$, (c) outer-target $T_e$, and (d) inner-target $T_e$. The inverse-cycle profiles remain close to both the forward prediction and SOLPS-ITER, demonstrating cycle consistency: parameters recovered by the inverse procedure reproduce the original fields.}
  \label{fig:nn_test_cases}
\end{figure*}

To complement the two-dimensional cycle-consistency results of Sec.~\ref{sec:results_cycle}, we illustrate the same forward/inverse round-trip at the level of the one-dimensional profiles extracted from the conditional U-Net. For a representative held-out test case (Fig.~\ref{fig:nn_test_cases}), the U-Net reproduces the SOLPS-ITER profiles with reasonable fidelity, recovering the correct amplitudes and characteristic profile shapes, both when evaluated at the true control parameters (forward) and when using the parameters recovered by the inverse procedure (inverse cycle). We emphasize that this forward/inverse comparison is a property of the U-Net surrogate of Sec.~\ref{sec:surrogate}; it is distinct from the query-by-committee profile ensemble of Sec.~\ref{sec:profile_ensemble}, which performs no parameter inversion.

\bibliography{paper}

\end{document}

%% file: table_metrics_paper_1.tex
\newcommand{\tabIImetricsbody}{%
  \rowcolor{blue!5}
  $T_e$ & eV & 12.24 & 1.26 & 0.994 & 0.996 \\
  $T_i$ & eV & 16.71 & 1.54 & 0.992 & 0.987 \\
  \rowcolor{blue!5}
  $n_e$ & m$^{-3}$ & $2.61 \times 10^{18}$ & 0.62 & 0.987 & 0.964 \\
  $u_a$ & m$\cdot$s$^{-1}$ & 940.9 & 2.54 & 0.975 & 0.976 \\
}